\documentclass[pdflatex,sn-nature]{sn-jnl}%

\usepackage{graphicx}%
\usepackage{multirow}%
\usepackage{amsmath,amssymb,amsfonts}%
\usepackage{amsthm}%
\usepackage{mathrsfs}%
\usepackage[title]{appendix}%
\usepackage{xcolor}%
\usepackage{textcomp}%
\usepackage{manyfoot}%
\usepackage{booktabs}%
\usepackage{algorithm}%
\usepackage{algorithmicx}%
\usepackage{algpseudocode}%
\usepackage{listings}%
\usepackage{subcaption}%
\usepackage{afterpage}
\usepackage{natbib}
\usepackage{tcolorbox}
\usepackage{tikz}
\usetikzlibrary{calc,intersections,patterns,arrows.meta,positioning}
\usepackage[myheadings]{fullpage}
\usepackage{standalone}
\usepackage{booktabs}
\usepackage{pgfplots}
\usepackage{lineno}

\pgfplotsset{compat=1.18}

\begin{document}

\title[Article Title]{Inferring Mechanisms from Bits: Promises and Pitfalls of Higher-Order Information in Cognitive Science}

\author*[1,2]{\fnm{Daniel} \sur{Rebbin}}\email{dlrebbin@gmail.com}

\author[2,3,4]{\fnm{Keenan J. A.} \sur{Down}}

\author[5,6]{\fnm{Thomas F.} \sur{Varley}}

\author[7]{\fnm{Robin A.A.} \sur{Ince}}

\author*[2,8,9]{\fnm{Andres} \sur{Canales-Johnson}}\email{afc37@cam.ac.uk}

\affil[1]{\orgdiv{Brain and Cognition Center}, \orgname{University of Amsterdam}, \orgaddress{\city{Amsterdam}, \country{NL}}}

\affil[2]{\orgdiv{Department of Psychology}, \orgname{University of Cambridge}, \orgaddress{\city{Cambridge}, \country{UK}}}

\affil[3]{\orgdiv{Department of Psychology}, \orgname{Queen Mary University of London}, \orgaddress{\city{London}, \country{UK}}}

\affil[4]{\orgdiv{Department of Computing}, \orgname{Imperial College London}, \orgaddress{\city{London}, \country{UK}}}

\affil[5]{\orgdiv{Vermont Complex Systems Institute}, \orgname{University of Vermont}, \orgaddress{\city{Burlington}, \country{USA}}}

\affil[6]{\orgdiv{Department of Psychological and Brain Sciences}, \orgname{Indiana University}, \orgaddress{\city{Bloomington}, \country{USA}}}

\affil[7]{\orgdiv{School of Psychology and Neuroscience}, \orgname{University of Glasgow}, \orgaddress{\city{Glasgow}, \country{UK}}}

\affil[8]{\orgdiv{CINPSI Neurocog, Faculty of Health Sciences}, \orgname{Universidad Católica del Maule}, \orgaddress{\city{Talca}, \country{CL}}}

\affil[9]{\orgdiv{Neuroscience Center, Helsinki Institute of Life Science}, \orgname{University of Helsinki}, \orgaddress{\city{Helsinki}, \country{FI}}}

\abstract{Higher-order information theory has become a rapidly growing toolkit in computational neuroscience, motivated by the idea that multivariate dependencies can reveal aspects of neural computation and communication invisible to pairwise analyses. Yet functional interpretations of synergy and redundancy often outpace principled arguments for how statistical quantities map onto mechanistic cognitive processes. Here, we review the main families of higher-order measures with the explicit goal of translating mathematical properties into defensible mechanistic inferences. We move from undirected through directed and dynamic metrics towards interventional measures, tracing an ascent toward increasingly causally relevant notions of information while making explicit the mathematical, computational, and interpretational costs incurred at each step. Equipped with the relevant mathematical and interpretational essentials, we connect the redundancy-synergy balance to cognitive function by progressively embedding these information-theoretic concepts within real-world constraints, starting with small synthetic systems and gradually building up to large-scale neuronal recordings. We argue that balanced layering of synergistic integration and redundant broadcasting optimizes multiscale complexity, formalizing a computation-communication tradeoff. We conclude by identifying three key directions for future mechanistic insight: cross-scale bridging, intervention-based validation, and thermodynamically grounded unification of information dynamics.}

\maketitle

\section{Introduction}\label{sec1}

Higher-order information theory has gained increasing popularity within computational neuroscience to investigate how the brain's many components share information to serve cognitive functioning. Information theory has been argued to be the \textit{lingua franca} for the modeling of complex systems \cite{Varley2025}, such as the brain. If we grant this, extending the vocabulary beyond its usual domain of bivariate interactions to three or more variables would represent a valuable step towards capturing higher-order cognitive functioning. Cognition is often characterized as an emergent property of the system \cite{Miller.etal2024}, where the whole is irreducible to the parts in isolation. Higher-order information theory is uniquely sensitive to emergent effects that support cognition and that conventional methods are insensitive to. Stereoscopic vision, for instance, is a canonical example of a synergistic effect \cite{Luppi.etal2022}: It is to be found in neither eye nor hemisphere in isolation but only through their triangulation with the scene information that they capture complementarily. In contrast, redundant information is information copied across multiple subsets of elements, such as color information from overlapping visual fields. A blind spot of established methods for studying such key cognitive function might be filled in by higher-order information theory. 

The alluring promise of synergy and redundancy to extend established methods to such emergent effects has inspired various functional interpretations in the cognitive neuroscience literature, from their role in multisensory feature integration \cite{Park.etal2018} to explaining psychedelic effects on consciousness \cite{Luppi.etal2024b}. These efforts have been mostly facilitated by the partial information decomposition (PID) \cite{Williams.Beer2010}, a proposal to solve the decades-old problem of understanding the negative values that occur when generalizing the notion of information sharing to three or more variables \citep{Shannon1948, McGill1954, Watanabe1960, James.etal2011}. 

That promise, however, risks outpacing its justification, echoing Shannon’s famous warning against the “bandwagon” of applying information theory beyond the domains for which its concepts had been carefully established \cite{Shannon1956}. It is tempting to ascribe functional readings of synergy and redundancy before it is clear how a statistical quantity computed from recorded activity licenses a claim about the underlying
cognitive process. Since its inception and subsequent tutorials \cite{Timme.etal2014} for applying higher-order information measurement tools, questions around fundamental theoretical disagreements (What does redundancy mean in mathematically precise terms?) and functional interpretation have become more immediate \cite{Varley2024a}. For this reason, we regard it as an appropriate moment to recapitulate the most essential theory to carefully build towards a well-founded functional interpretation when measuring higher-order information. To avoid model reification and interpretational pitfalls, we argue to distinguish mathematical `map' of our models and functional `territory' of the cognitive systems they model \cite{Andrews2021,Hoel2017}. Higher-order statistical behavior and (higher-order) generative processes need to be clearly distinguished \cite{Rosas.etal2022}, with careful inference required to bridge this gap.

\begin{figure}[H]
\centering
\includegraphics[width=1\linewidth]{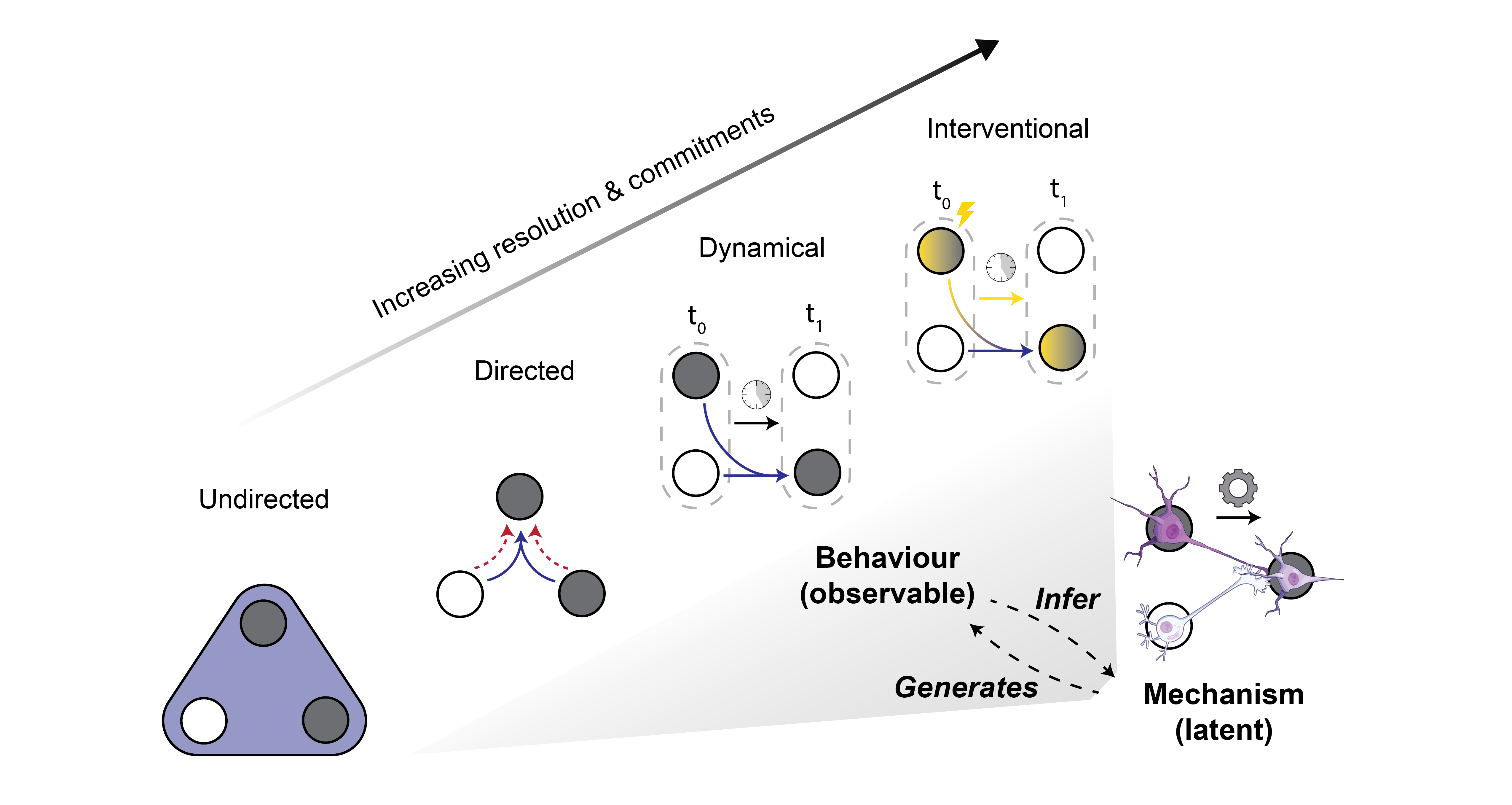}
\caption{Each family of higher-order information measures answers a sharper question about a system's higher-order structure at a correspondingly greater experimental, mathematical, and interpretational cost. \textit{Undirected} measures (e.g.\ O- and S-information) ask only whether higher-order structure is present, treating all variables symmetrically. \textit{Directed} decompositions (PID) ask which variables carry it, imposing a source-target split (red: redundant; blue: synergistic/unique contributions). \textit{Dynamical} measures (e.g.\ transfer entropy, integrated information decomposition) let the arrow of time fix that split, asking how structure propagates from one moment ($t_0$) to the next ($t_1$). \textit{Interventional} measures ask how it responds to a perturbation actively set on a node rather than passively observed, moving a step closer towards causal inference. All four families are sampled from recorded activity and therefore yield a higher-order \textit{behaviour}: a statistical signature of the observed distribution. The generative \textit{mechanism} that produced that distribution is latent and never measured directly. Crucially, the four rungs are not steps toward the functional mechanism but successively finer characterisations of behaviour; the step from behaviour to mechanism is a distinct inferential leap (dashed arrows). A mechanism generates observable behaviour, whereas recovering it requires \textit{inferring} across a many-to-one map, the central challenge this review examines.}
\label{overview}
\end{figure}

The central purpose of this review is to examine that inferential
step directly. We organise the available measures as a ladder of increasing
resolution (Figure \ref{overview}): undirected measures that ask only to what extent higher-order information is present; directed decompositions that ask which variables carry it; dynamic measures that ask how it evolves in time; and interventional measures ask how it responds to manipulation. Every step upward exacts heavier experimental, mathematical, and interpretational commitments, a tradeoff we want to make explicit. By starting with simple models, we emphasize intuitive understanding that can inform sound application. Concretely, we seek to connect the purely mathematical properties of higher-order information measures to their capacities and limitations in furnishing mechanistic explanations. We first systematize entropy-based generalisations of mutual information and contrast them with information decompositions, which represent an attempt to overcome their limitations. The overarching goal is pragmatic: to equip the experimentalist with a principled understanding of what each measure can and cannot reveal about neural function, and how to deploy it to advance mechanistic insight. 

\section{Tiers of Complexity: From Shannon Entropy to Higher-order Decompositions}

\subsection{Undirected Higher-order Information: Diverging Generalisations of Mutual Information}

\begin{figure}[h]
\centering
\includegraphics[width=1\linewidth]{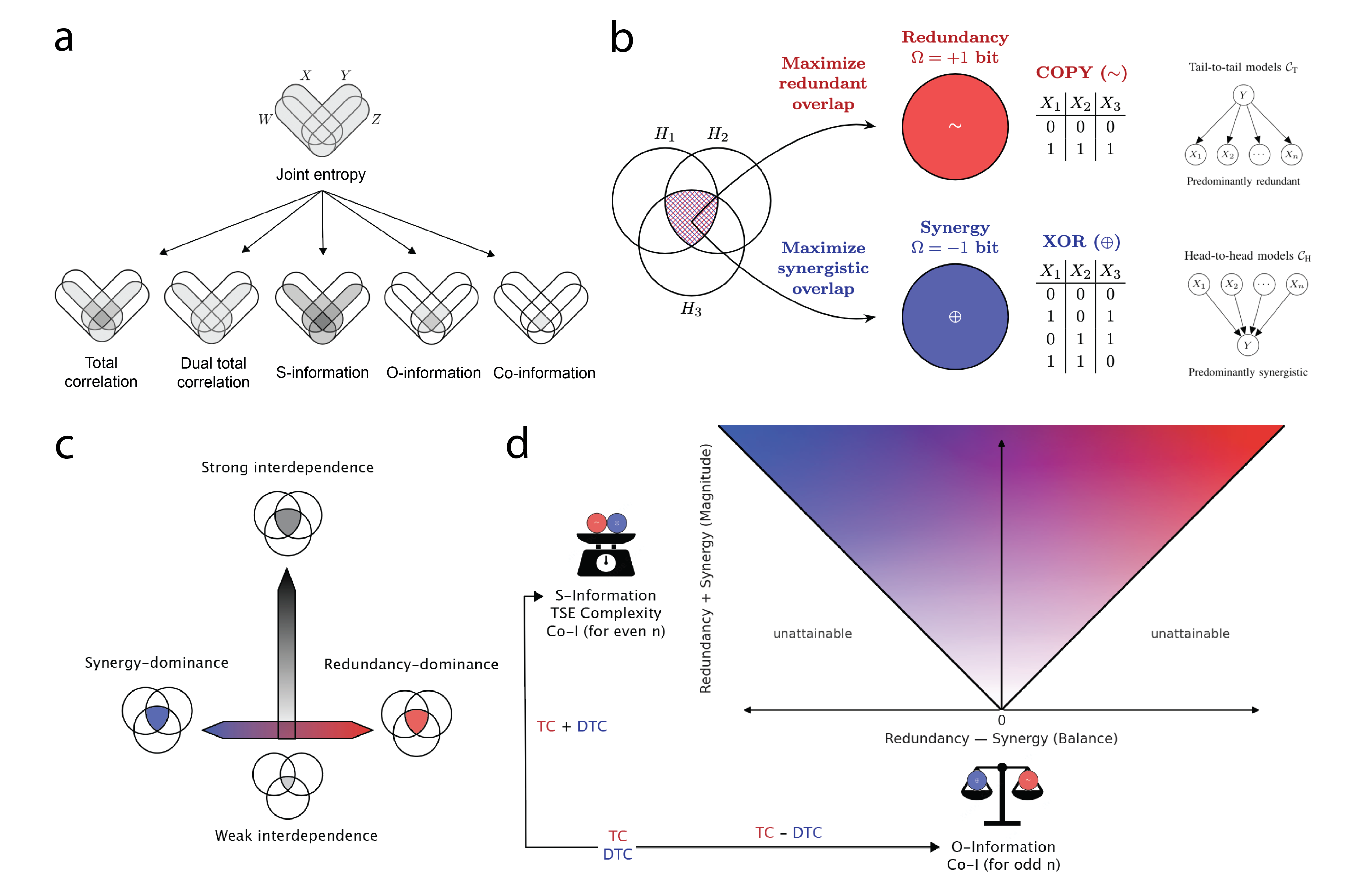}
\caption{a) The mutual information $I(X_1;X_2)$ between two random variables $X_1$ and $X_2$ can be thought of as corresponding to the \textit{shared randomness} of the joint entropy (shaded in grey). Beyond two variables, various metrics account for the notion of information sharing in different mathematical ways. b) On the left, redundancy and synergy are shown overlapping at the intersection of the shared entropy between three variables. As the circles are rotated onto another, two qualitatively different, yet dual, prototypes of higher-order interaction can emerge: In the top row, we see the only purely redundant dependency, signified by a COPY distribution in red. As shown in the accompanying truth table, all states are perfectly synchronized, acting as a single `giant bit'. The XOR gate's synergistic interdependence also constrains the parts, but each variable is determined only by the joint state of the other two, not by either alone. The associated directed graphs (rightmost figures) show a duality between the broadcasting `tail-to-tail' model (redundancy) and the integrative `head-to-head' models (from \cite{Rosas.etal2024}): redundancy allows for the `deduction' of all other states from a single one, while synergy requires all other states to `induce' the state of any one of them. c) Higher-order information sharing can vary along two axes: (1) the vertical dimension represents the overall strength of the interdependence (opacity gradient), increasing with greater entropy overlap, and (2) the extent to which synergy or redundancy dominates (blue-red gradient), a measure of balance on the horizontal axis. d) The relationship between the different higher-order metrics reviewed is related to the (bounded) coordinate system that results from c). Importantly, a system cannot be more redundant or synergistic than it is interdependent overall (signified by the `unattainable' regions for which $|x| > y$).}
\label{undirected}
\end{figure}

Shannon's original work \cite{Shannon1948} was concerned with the engineering task of lossless transmission of information from a sender A to a receiver B. Still, the metric of information he created is itself undirected. Instead of capturing a physical, causally powerful flow of information, it has instead been shown to be much closer to a generalized notion of correlation between two variables that can subsume various other linear, bivariate tests of statistical dependence \cite{Ince.etal2017}. Shannon's \textbf{entropy} quantifies the uncertainty in a variable $X$,
\begin{align}
    H(X) = -\sum_{x} p(x)\log_2 p(x),
\end{align}
measured in bits. Choosing, by convention, the base 2 for our logarithm, this can be conceived of as how often the space of possible outcomes must be halved to specify a particular outcome uniquely \citep{Wheeler2018}. The logarithm translates this multiplicative process into an additive language, where the primary unit of study is the bit. If, for instance, a neuron fires as often as it stays silent ($p(x = 1) = p(x = 0) = 0.5$), its entropy $H(X)=1$ bit, since we could not be more uncertain about what event we would subsequently observe. The conditional entropy $H(X_1 \mid X_2)$ is the uncertainty that remains in $X_1$ once $X_2$ is known. Two neurons firing in perfect (a)synchrony, each with 1 bit of entropy individually but just one bit of \textbf{\textit{joint} entropy} ($H(X_1, X_2) = 1$), would then have 0 conditional entropy ($H(X_1 \mid X_2) = H(X_2 \mid X_1) = 0$). \textbf{Mutual information} (MI, or, simply $I$) follows from entropy as the reduction in uncertainty about one variable afforded by the other, which would be maximal for our two perfectly dependent neurons. It can be written in three algebraically equivalent ways:
\begin{align}
    I(X_1;X_2)
    &= H(X_1) + H(X_2) - H(X_1,X_2) \label{eq:mi-sum}\\
    &= H(X_1,X_2) - \big[H(X_1\mid X_2) + H(X_2\mid X_1)\big] \label{eq:mi-dual}\\
    &= D_{\mathrm{KL}}\!\big(P(X_1,X_2)\,\|\,P(X_1)P(X_2)\big). \label{eq:mi-kl}
\end{align}
For two variables these yield the same number, the shared randomness of the joint entropy. For three or more variables, though, their different algebraic approaches to calculating dependency come apart, and each equivalent form generalises to a different multivariate measure and associated quantity. Although their different ways of probing dependency coincide mathematically for $n=2$, for our inquiry into their interpretation, it is helpful to think of them as effectively querying the system in different ways (Figure \ref{undirected} a). We may ask:

\begin{enumerate}
    \item To what extent are the parts coupled (equation \ref{eq:mi-kl})? (\textbf{Total Correlation})
    \item To what extent is the whole constrained by relationships among the parts (equation \ref{eq:mi-dual})? (\textbf{Dual Total Correlation})
    \item How much information is shared between all variables (equation \ref{eq:mi-sum})? (\textbf{Co-information})
    \item To what extent does redundancy dominate synergy? (\textbf{O-information})
    \item What is the aggregate information sharing strength? (\textbf{S-information})
    \item How much information crosses bipartitions across scales? (\textbf{TSE complexity})
\end{enumerate}

We can illustrate what the (quantitative) answers to these questions (Table \ref{tab:measures_xor_copy}) mean best by considering the systems that embody the extremes of redundancy and synergy: the COPY (`giant bit') and XOR (`parity') distributions (Figure \ref{undirected} b). They represent the prototypical representations of what we actually mean when we speak about redundancy and synergy (indeed, the XOR gate is the only system which, in all scenarios, exhibits synergy \cite{Down.Mediano2025}). While each part tells us everything about the whole in a redundant system, nothing short of the whole is necessary to define the state of each part in a synergistic one. Analogously, the joint entropy of the whole $H(X_1, X_2, X_3) = 1$ bit for the COPY distribution, $H(X_1, X_2, X_3) = 2$ bits for the XOR distribution. Although each variable, considered individually, is equally likely to be 0 or 1 for both, the joint state introduces additional uncertainty in the synergistic case. As we will now explore, different queries to the system account for these intrinsically higher-order dependencies in various ways.

\begin{table}[ht]
\centering
\caption{Comparison of six information measures for XOR, COPY \& AND distributions ($n=3$, fair binary variables). Values in bits.}
\begin{tabular}{lccl}
\toprule
Measure & XOR & COPY  &AND\\
\midrule
Total correlation, $TC$ & 1 & 2  &0.811\\
Dual total correlation, $DTC$ & 2 & 1  &1\\
Co-information, $\mathrm{Co\textendash I}$ & $-1$ & $1$  &$-0.189$\\
O-information, $\Omega=TC-DTC$ & $-1$ & $1$  &$-0.189$\\
S-information, $\Sigma=TC+DTC$ & 3 & 3  &1.811\\
TSE complexity, $C_{TSE}$ & 1 & 1  &0.604\\
\bottomrule
\end{tabular}
\label{tab:measures_xor_copy}
\end{table}

\subsubsection{Total correlation: To what extent are the parts coupled?}
From the statistician's perspective, the most salient application of information theory is testing for dependence between variables. Indeed, mutual information can provide a versatile nonparametric measure that can substitute for a whole array of statistical tests (ANOVA, chi-squared, GLM, and decoder fit, etc.) \cite{Ince.etal2017}. This frames information as evidence towards deviation from independence, as quantified by the KL-divergence. Conceiving of information as such a generalized test against independence (equation \ref{eq:mi-kl}) generalises to the \textbf{total correlation} (TC):
\begin{equation}
    TC(\textbf{X}) = D_{\mathrm{KL}}(P(\textbf{X}) || \prod_i P(X_i)) = \left( \sum_{i=1}^N H(X_i) \right) - H(\textbf{X})
\end{equation}
A higher TC value indicates more evidence for a model that characterizes a system’s components as dependent, since a small entropy of the joint state $H(\textbf{X})$ detracts little from the sum of individual contributions. The bias towards redundant coupling leads to a smaller joint entropy and, therefore, higher TC for the COPY than for the XOR distribution. Thus, the closer the joint distributions approach a single, synchronized system, the higher the TC.

\subsubsection{Dual total correlation: To what extent is the whole constrained by relationships among the parts?}

While the TC has a bias towards lower-order terms in its notion of information sharing, its dual metric, the \textbf{dual total correlation} (DTC), also called binding information \citep{Abdallah.Plumbley2012}, has recently been found to lean more towards higher-order terms \cite{Rosas.etal2025}. It derives from writing the MI as a subtraction of conditional entropy from the joint state ( equation \ref{eq:mi-dual}) and generalizing this algebraic approach to higher dimensions yields
\begin{equation}\label{DTC}
    \text{DTC}({\textbf{X}}) = H(\textbf{X}) - \left( \sum_{i=1}^N H(X_i|X_{-i}) \right),
\end{equation}
where conditioning each $X_i$ on all other variables $X_{-i}$ removes the non-shared entropy of each variable, also called the \textbf{residual entropy} \citep{James.etal2011}. Since the implicit question posed by applying the DTC emphasizes that the individual entropies cannot explain away the entropy of the whole, the bit values are flipped relative to the TC between XOR and COPY: the XOR distribution carries more information in its joint state and therefore has a higher DTC. 

\subsubsection{Co-information: How much information is shared between all variables?}

The \textbf{Co-information} (Co-I), also called interaction information \citep{McGill1954} (differing only by sign convention), derives from writing information as a sum of joint entropy terms (equation \ref{eq:mi-sum}, where the sign alternates over set sizes to avoid double-counting (the inclusion-exclusion principle):
\begin{equation}
    \mathrm{Co\textendash I}(X_1, …, X_N) = \sum_{\emptyset\neq S\subseteq\{X_1,\ldots,X_N\}}
(-1)^{|S|+1}H(S)
\end{equation}
In this way, the Co-I measures only the dependency present in \textit{all} variables considered, with non-zero (and possibly negative) values indicating higher-order dependencies. Unlike the TC and DTC metrics, Co-I can yield negative information for $n \geq 3 $, which was only interpreted as synergy-dominance much later. 

\subsubsection{O-information: To what extent does redundancy dominate synergy?}

We arrive at a particularly popular method for describing the redundancy-synergy \textit{balance}: the \textbf{O-information} ($\Omega$) \citep{Rosas.etal2019}. After originally naming it the enigmatic information, due to its then unclear interpretation, \citep{James.etal2011} showed that the O-information (organisational information) can be written as

\begin{equation}\label{O-information}
    \Omega(\textbf{X}) = \text{TC}(\textbf{X}) - \text{DTC}(\textbf{X}),
\end{equation}

allowing for the not-so-enigmatic interpretation as the overall tendency of a system to be synergy-dominated (for negative values) or redundancy-dominated (for positive values). Although TC and DTC also exhibit biases towards measuring redundancy or synergy, respectively, they conflate their contributions with the overall information-sharing \textit{strength}, which does not distinguish between synergy and redundancy and also includes bivariate dependencies (cf. Figure \ref{undirected}) \cite{Rosas.etal2025}. When one subtracts the synergy-associated DTC from the redundancy-associated TC, we cancel out any notion of information sharing \textit{strength} implicit in them and leave only their association with redundancy and synergy. The O-information gives a strictly higher-order account of information sharing in the sense that it is only sensitive to interactions between at least three variables (compare with the Venn diagram in Fig. \ref{undirected}a). If and only if $n=3$, these purely higher-order interactions live at the intersection between all variables, therefore coinciding then with the Co-information. 

\subsubsection{S-information: What is the aggregated information sharing strength?}

Complementary to the question whether redundancy or synergy dominates (O-information) stands the \textbf{S-information} ($\Sigma$) \citep{James.etal2011}, which sums (hence, \textit{S}-information) both synergy and redundancy together to give an easily computable overview of the information strength within the system. By summing up the mutual information between each variable and the rest of the system, we get:
\begin{equation}\label{S-equals-TC-plus-DTC}
\Sigma(\mathbf{X}) = \mathrm{TC}(\mathbf{X}) + \mathrm{DTC}(\mathbf{X}) = \sum_i I(X_i;X_{-i}).
\end{equation}
Being agnostic about low- or high-order contributions to the dependency, it yields the same results for the COPY and XOR distributions, where each of the three variables shares one bit with the rest, albeit in different ways that the S-information is insensitive to. 

\subsubsection{TSE complexity: How much information crosses bipartitions across scales?}

While the S-information indicates how much information \textit{single} parts share on aggregate with the rest of the system, the Tononi-Sporns-Edelman complexity, or short, \textbf{TSE complexity} ($C_{\mathrm{TSE}}$) \citep{Tononi.etal1994}, instead provides the average information over all \textit{sets} of variables $a$ with the rest (bipartitions): 
\begin{equation}\label{TSE-MI}
C_{\mathrm{TSE}}(\mathbf{X})
= \frac{1}{2}\sum_{k=1}^{N-1}
\binom{N}{k}^{-1}\sum_{\substack{|a|=k}} I(X_a;X_{-a}),
\end{equation}
where the factor 1/2 prevents the same bipartitions from being counted twice. Originally introduced to simultaneously capture segregated and integrated system dependencies \cite{Tononi.Sporns2003}, TSE complexity does not necessarily vanish under global synchronization. For example, if $X_1=\cdots=X_N=Z$ for a non-degenerate discrete random variable $Z$, then $C_{\mathrm{TSE}}=(N-1)H(Z)/2$. This is precisely what we see reflected in the COPY distribution example, where $C_{\mathrm{TSE}}$ is equal to 1 bit (instead of 0). $C_{\mathrm{TSE}}$ should therefore be interpreted as quantifying dependence distributed across subsystem scales, rather than as intrinsically distinguishing structured multiscale integration from redundant common-mode dependence.

\subsection{Comparing the Multivariate Generalisations of Mutual Information}

When contrasting these various metrics, all can be argued to be `natural' extensions of MI, as they define information sharing in different ways. There is only one way to characterize information sharing between two variables: by mutual information, a one-dimensional scalar. One might suppose that the number of components characterizing information sharing in higher dimensions would scale with the number of a system's elements. Regardless of the number of variables $n$, though, it has been recently found to be fully captured by two components: information sharing \textit{strength} and a second axis of \textit{balance} between redundancy and synergy \citep{Rosas.etal2025} (Figure \ref{undirected}c).

Although the two dimensions of information strength and balance can vary independently, they still share an important link through the roles redundancy and synergy play in both. While the O-information and the co-information (for odd \textit{n}) weigh them \textit{against} one another, like an analogue scale, the S-information and TSE complexity weigh them stacked on top of one another (together with the exclusively bivariate interactions), like a digital scale (Figure \ref{undirected}d). Fittingly, the former are called skew-symmetric and the latter symmetric measures of information sharing \citep{Rosas.etal2024}. The XOR and COPY distributions are extreme examples in different ways: They provide the maximal information-sharing strength for a three-variable system and the maximal contrast in redundancy-synergy tilt, just as the most extreme dissociation between these two (symmetric and skew-symmetric) dimensions. 

From a practical perspective, using the S-information and the O-information in concert provides two values that approximately capture (a) how strongly information is shared and (b) whether the whole or the parts predominantly hold it with minimal computational overhead. Generally, the co-information, total correlation and dual total correlation do not fall neatly into these two categories. The TSE complexity scales much more poorly than the S-information, while correlating strongly with it in neuroscientific data \cite{Varley.etal2023a}. Hence, the S- and O-information provide a solid initial overview of higher-order dependencies, delivering cheaply scalable and complementary insights into information sharing \textit{strength} as well \textit{balance} in a multivariate system. 

Overall, undirected entropy-based measures have established themselves as valuable tools that require few assumptions beyond the core tenets of information theory and can quickly provide answers to key questions about large systems, such as neural ensembles. However, neural systems usually have an asymmetry between inputs and outputs, unlike the XOR and COPY distributions, that we might want to capture in our analysis. The AND gate represents here a simple example with such an input-output asymmetry: the outcome variable $X_3$ is only in state 1 if both inputs $X_1$ and $X_2$ are both in state 1 as well and 0 otherwise. This can be compared to a neuron that is only pushed past threshold when two afferent inputs introduce synaptic drive simultaneously. 

When comparing the bit values for the different measures applied to the AND gate (Table \ref{tab:measures_xor_copy}), we immediately see that they all land somewhere between those of the XOR and COPY distribution. The overall information \textit{strength} is not as pronounced: On average 0.604 bits (instead of 1 bit for COPY and XOR) are shared across a cut through the system ($C_{TSE}$), summing over three variables to 1.811 bits (instead of 3 bits) on aggregate ($\Sigma$). This intermediate \textit{strength} of information sharing for the AND gate exhibits a slightly tilted \textit{balance} towards synergy over redundancy ($\Omega = -0.189$ bits); Extra information beyond bivariate coupling emerges from taking the \textit{context} (conditioning on the third variable in this case) into account: the hallmark of synergy. Hence, it is interesting to note that our idealised three-neuron system's AND-gate-like thresholding of summed presynaptic activity performs a computation that is more synergistic than redundant. 

As we can see from these examples, undirected metrics use nothing but Shannon's core tenets of information theory to probe higher-order dependencies in various ways that Shannon's mutual information is mute on. The fact that a single bit value is the response to each (implicit) query is maximally concise, but therefore also lowest-resolution. One might ask two related questions about these higher-order dependencies to probe them more in-depth: (1) How much synergy and redundancy are there within the system exactly, and (2) how are they distributed over sets of variables? In the case of the XOR and COPY distributions, the answer seems straightforward: each variable pair holds one bit of synergistic and redundant information, respectively, about the third. Synergy and redundancy are clearly dissociable in their diametrically opposed representations. However, for the AND gate, where some degree of both is involved, the answers are not nearly as straightforward as they might first appear.

\subsection{Directed Higher-Order Information: Partial Information Decomposition(s)}

\begin{figure}[t]
\centering
\includegraphics[width=1\linewidth]{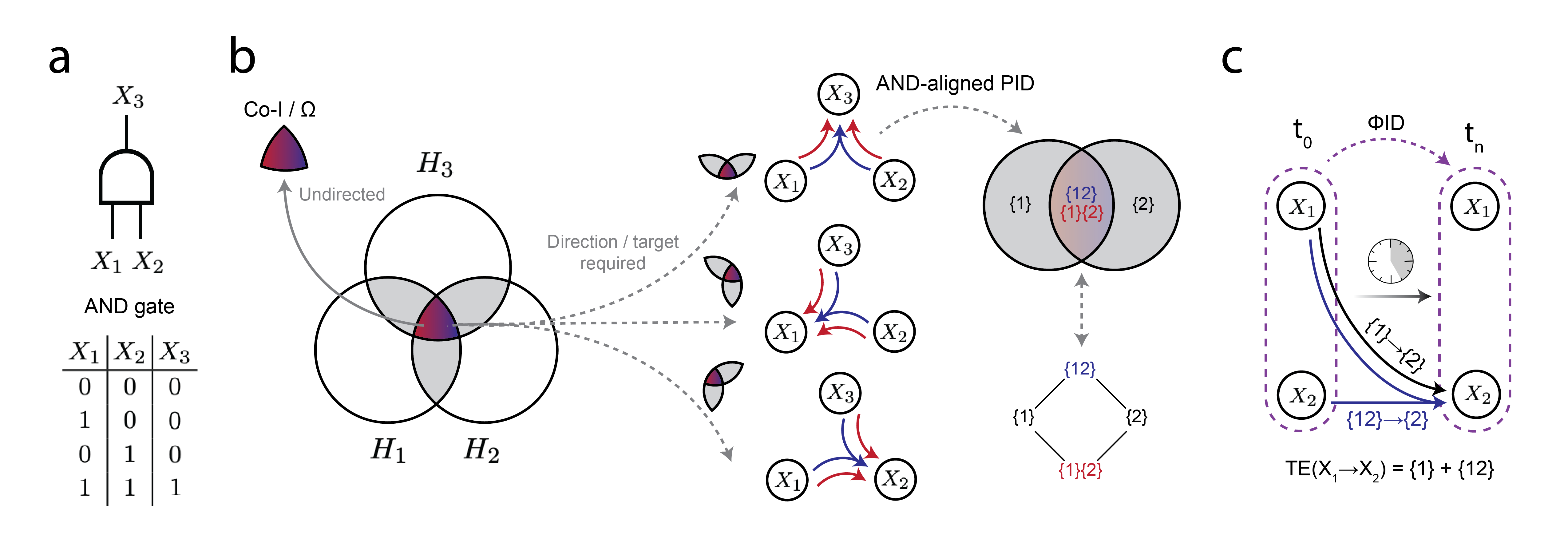}
\caption{a) Unlike XOR or COPY, the AND gate provides an impure redundancy-synergy mix and an asymmetry in its truth table that suggests a natural target variable. b) Depending on the choice of which variables are defined as sources and which as targets, the three-variable case admits three distinct decompositions of the underlying shared entropy, whereas Co-I and O-information are only concerned with the triple intersection. This yields one PID `aligned' with the input-output flow (and two `misaligned' PIDs), where the information that the two input variables $X_1$ and $X_2$ hold individually about the target variable $X_3$ can be decomposed into unique ($\{1\}$ and $\{2\}$), redundant ($\{1\}\{2\}$ in red) and synergistic ($\{12\}$ in blue) partial information atoms. c) Dynamical decompositions of information, such as $\Phi$ID, allow for insights into how elements translate information from the past onto the future, showing that the widely used TE metric not only contains the transfer of unique information from source to target but also synergy between the source's and target's joint past.}
\label{directed}
\end{figure}

Up to this point, the lack of insight into how information is exactly distributed over variables to compose undirected higher-order metrics and how to interpret negative (co-)information values lingered as problems that are as much mathematical as interpretational. In their seminal paper, Williams and Beer \citep{Williams.Beer2010} targeted the shortcomings of existing multivariate generalizations of mutual information, as reviewed above, and proposed what became the cornerstone of a new mathematical program, called the Partial Information Decomposition (PID): the non-negative decomposition of information shared between three or more variables. The authors posited that the information multiple \textit{source} variables hold about a \textit{target} variable (separated by a semicolon, such as in $I(X_1,X_2;X_3)$) can be decomposed into unique, redundant, and synergistic `atoms' (Figure \ref{directed} b). These atoms are the smallest elements that jointly compose the joint mutual information. In contrast to the measures previously reviewed, which are undirected, directionality is here baked into its approach, so that we consistently quantify information \textit{about} the target in particular. We see in Figure \ref{directed} b, that this leads to \textit{n} different decompositions for \textit{n} variables. If we take $X_3$ to be the target for $n=3$, then we write:

\begin{align}
    I(X_1,X_2;X_3) = I_{\partial}(\{1\}\{2\}) + I_{\partial}(\{1\}) + I_{\partial}(\{2\}) + I_{\partial}(\{12\}),
\end{align}

where $I_{\partial}(\{1\}\{2\})$ is the atom representing the information in either $X_1$ or $X_2$ (redundancy) about $X_3$, $I_{\partial}(\{12\})$ is the information about $X_3$ only available by considering $X_1$ and $X_2$ jointly (synergy), while $I_{\partial}(\{1\})$ and $I_{\partial}(\{2\})$ represent the information unique to each source. The decomposition can be algebraically unlocked by defining a function that describes one of the atoms, usually the redundant one, which then allows assigning bit values to all synergistic and unique information atoms.

Despite the promise of cutting multivariate information at its joints in this way, there are three main issues with the PID that we want to highlight: (1) Finding a redundancy function, or what axioms it should even satisfy, turns out to be more troublesome than expected, (2) the inherent distinction between sources and targets brings additional, contentious assumptions with it, and (3) its computational as well as interpretational complexity that constrain its practical usefulness.

To illustrate issues (1) and (2), we consider the AND gate again (see Figure \ref{directed}). In this case, we \textit{do} have a designated target variable, an asymmetry in the truth table that was absent for the XOR and COPY distributions. This might or might not \textit{align} with the source-target distinction that the PID forces us to make, effectively yielding one `aligned' and two `misaligned' decompositions. In other words, `alignment' here means that the target in the PID is actually the output variable of the AND gate. Looking at Table \ref{tab:and_pid_aligned_misaligned}, we can see that a subset of redundancy functions disagree in their assignment of bit values (1) to unique, redundant, and synergistic atoms and (2) when comparing `aligned' and `misaligned' target variable choices. 

\begin{table}[ht]
\centering
\caption{Selected PID atoms (bits) for the equiprobable AND gate under several redundancy functions. 
Aligned decomposition use sources $(X_1,X_2)$ and target $X_3=X_1 \wedge X_2$. 
Misaligned decompositions use sources $(X_1,X_3)$ and target $X_2$ or sources $(X_2,X_3)$ and target $X_1$.}
\label{tab:and_pid_aligned_misaligned}
\begin{tabular}{lcccccccc}
\toprule
& \multicolumn{4}{c}{Aligned} 
& \multicolumn{4}{c}{Misaligned} \\
\cmidrule(lr){2-5} \cmidrule(lr){6-9}
Redundancy function
& $R$ & $U_0$ & $U_1$ & $S$
& $R$ & $U_0$ & $U_1$ & $S$ \\
\midrule
$I_{\min}$       & 0.311 & 0.000 & 0.000 & 0.500 & 0.000 & 0.000 & 0.311 & 0.189 \\
$I_{\mathrm{mmi}}$ 
                 & 0.311 & 0.000 & 0.000 & 0.500 & 0.000 & 0.000 & 0.311 & 0.189 \\
$I_{\mathrm{BROJA}}$ 
                 & 0.311 & 0.000 & 0.000 & 0.500 & 0.000 & 0.000 & 0.311 & 0.189 \\
$I_{\mathrm{proj}}$ 
                 & 0.311 & 0.000 & 0.000 & 0.500 & 0.000 & 0.000 & 0.311 & 0.189 \\
$I_{\mathrm{ccs}}$ 
                 & 0.104 & 0.208 & 0.208 & 0.292 & 0.000 & 0.000 & 0.311 & 0.189 \\
$I_{\pm}$& 0.561 & -0.250 & -0.250 & 0.750 & 0.061 & -0.061 & 0.250 & 0.250 \\
$I_{sx}$         & 0.123 & 0.189 & 0.189 & 0.311 & 0.000 & 0.000 & 0.311 & 0.189 \\
\bottomrule
\end{tabular}
\end{table}

First, the redundancy functions disagree on redundancy and synergy individually but agree exactly on their difference, recovering the O-information ($\Omega = -0.189$, as in Table \ref{tab:measures_xor_copy}) of the AND gate. $I_{\pm}$, $I_{sx}$, and $I_{ccs}$ show mutually incompatible results from the remaining shown redundancy functions that at least show consistency in this simple case of the AND gate (of which we have collected only seven popular choices - see \cite{Liardi.etal2026} for an exhaustive overview). Worryingly, the redundancy disagree more strongly when the choice of the target variable was aligned with the logic flow from inputs to outputs for the AND gate. When applied to neuroscientific data, $I_{\pm}$ and $I_{sx}$ ($I_{\pm}$'s differentiable cousin) deviate again strongly from the others in their results, since they permit negative atom values (together with $I_{ccs}$). This contradicts the original purpose of a \textit{non-negative} decomposition \cite{Williams.Beer2010} but is argued for as a desirable property of indicating misinformation about a target variable \cite{Finn.Lizier2018a, Ince2017a}. Also for the empirical data, several among the proposals have been shown to correlate strongly with each other \citep{Rosas.etal2020a}, or at least permit qualitatively similar conclusions, even when exact bit values differ \cite{Kay.etal2022}. However, one can hardly say that these conclusions are robust, though, since the results \textit{do} depend on the assumptions one has to commit to in choosing a redundancy function.

As more and more redundancy functions have been proposed \citep{Harder.etal2013, Bertschinger.etal2014, Lizier.etal2018, Ince2017a, Makkeh.etal2021, James.etal2018, Rosas.etal2020a} with, as yet, only meager convergence of opinions, a central problem at the core of the PID undertaking has become clear: Not only is redundancy taken to mean different things by different authors in their formalizations, but also has the inclusion of one set of axioms resulted in tradeoffs with the violation of other axioms deemed essential by other theorists. It has been proven that some axiom sets are irreconcilable with one another and require non-trivial mathematical commitments, especially when going beyond three variables \citep{Matthias.etal2025, Liardi.etal2026}. 

What adds to the extra mathematical commitments required for $n > 3$ is the superexponential growth in the number of PID atoms with $n$. Moving from two to four source variables already implies 166 (instead of four) distinct atoms, increasing to 7,828,352 for six sources (the $n$-th Dedekind number - 2), quickly becoming infeasible even for high-performance computing beyond that, whereas the family of entropy-based measures can scale easily to hundreds of variables. The substantial computational cost arising from the algorithmic mission of the PID is sweeping, aiming for a \textit{complete} decomposition rather than just a summary metric, such as the O-information, which only indicates overall tendencies toward redundancy or synergy. It seems neither feasible nor necessarily beneficial from an epistemological standpoint to aim for complete decompositions of larger systems, which quickly blow up in both computational and semantic complexity. 

From a pragmatic perspective, one can ameliorate PID's limitations in several ways. First, considering only triads (two sources and one target), as is usually done in practice, sidesteps some axiomatic and interpretational difficulties. Secondly, one should ensure that the inherent source-target directionality in the PID aligns with the causal source-target directionality in the system modeled, if a causal explanation is attempted. The previously mentioned study by \cite{Kay.etal2022} did this at a single-neuron level, for instance, by taking basal and apical dendrites as source and axonal action potential as target variables. Here, physiological mechanisms determine the target choice that PID requires \textit{a priori}. Usually, though, the causal directionality is not this clear from the outset, requiring a non-arbitrary choice of source and target variables, as demonstrated with the AND gate. PID requires a principled choice of the target variable that does not fall out of the static probability distribution it acts on. 

\subsection{Information Dynamics: From Random Variables to Random Processes}

Such a principle becomes available as soon as we stop treating our observations as random variables and start treating them as random processes. The future cannot influence the past, so temporal ordering imposes a source-target ordering that reflects an asymmetry that is a property of the world rather than of our analysis. Two routes follow. One sidesteps the target problem, tracking how interdependence evolves without nominating a target at all; the other exploits it, letting time fix sources and targets. One caveat governs both: the asymmetry time supplies is \textit{predictive} in the Wiener--Granger sense, licensing statements about which past states help forecast which future ones. It says nothing about what would happen were we to set a variable to a given value.
 
The first route extends the undirected measures of Section 2 by replacing information between variables with information \textit{rates} between processes \citep{Mijatovic.etal2021}. The O-information admits both a dynamic form tracking the redundancy--synergy balance as it evolves \citep{Stramaglia.etal2021} and a rate defined on processes with temporal correlations \citep{Faes.etal2022}, which matters because genuinely dynamical forms of redundancy and synergy can be misread when static measures are applied to temporally correlated data. Crucially, these add temporality but not directionality: the O-information rate remains symmetric under permutation of elements. For the same reasons given in Section 2 (rigorous grounding in Shannon entropy and scalability) we recommend these as an initial screen for large ensembles.

The second route yields the most widely used directed measure in the field. Transfer entropy (TE) quantifies how much the past of a source $X_2$ tells us about the future of a target $X_1$ over and above the target's own past \citep{Schreiber2000, Shorten.etal2025}:
 
\begin{align}
    \mathrm{TE}_{X_2 \to X_1} = I\!\left(X_1^{t+1} ;\,X_2^{0:t} \mid X_1^{0:t}\right).
\end{align}
 
Naively, one might think that this represents the information transferred from $X_2$'s past onto $X_1$'s future state, discounting the past influence (from time index $0$ to $t$) of the target variable $X_1$ on its own future. We might therefore believe that this corresponds to a flow of \textit{unique} information from $X_2$ to $X_1$. However, we can show with the PID that conditioning also leaves the \textit{synergistic} information between the shared past of $X_1$ and $X_2$ \cite{James.etal2016} (Figure \ref{directed}). When designating them as source variables, we can show that
 
\begin{align}
    \mathrm{TE}_{X_2 \to X_1} = I_{\partial}(\{2\}) + I_{\partial}(\{12\}). \label{TE_PID}
\end{align}
 
Therefore, TE conflates what the source's past uniquely carries with the synergy between the source's past and the target's own history and cannot distinguish them. This highlights that, while TE is a useful tool for the dynamic relationship between two variables, higher-order metrics are needed to capture emergent interactions unfolding in time \cite{Daube.etal2022}. Non-dynamical PID itself has been shown to be insufficient in capturing the transition from instantaneous to time-lagged interactions under Gaussian assumptions (such as applies to fMRI analyses, for instance); In that case, partial information \textit{rate} decomposition resolves unique, redundant and synergistic contributions per unit time rather than per observation \cite{Faes.etal2025} to isolate genuinely dynamical redundancy and synergy otherwise mischaracterised. 

A different approach to a PID of temporal processes, the Integrated Information Decomposition ($\Phi$-ID) \cite{Mediano.etal2019}, was introduced to extend the PID not only in time but also from many-to-one to many-to-many interactions. It concerns itself with the future-directed conversion of all partial information atoms from $t_0$ to $t_1$. This implies that we map all possible source and target sets onto one another. Single-target atoms of PID-in-time would be subsumed, such as $I_{\partial}(\{2\}\rightarrow\{1\})$ and $I_{\partial}(\{12\}\rightarrow\{1\})$ for the TE decomposition (compare with equation \ref{TE_PID}), with the latter being called `downward causation'. Crucially, new source-target combinations, like $I_{\partial}(\{12\}\rightarrow\{12\})$ are introduced. $I_{\partial}(\{12\}\rightarrow\{12\})$, for instance, describes the information that remains in the `whole' only, dubbed `causal decoupling', and would go undetected by a time-lagged PID due to its focus on single targets only. 

The increased resolution of information atoms that $\Phi$-ID aims for is certainly appealing to demonstrate precisely how information flows decompose over time and has mainly found application in large-scale human fMRI studies \cite{Luppi.etal2024d, Luppi.etal2023b}, macaque electrocorticography (ECoG) \cite{Rosas.etal2020} and at the level of neuronal circuits \textit{in vitro} \cite{Menesse.etal2024, Varley2023b}. However, the aforementioned shortcomings of PID are compounded rather than surpassed: Running $\Phi$-ID on merely three variables requires pairing all 18 past and 18 future PID atoms with one another, hence yielding $18^2=324 $ for $n=3$ and 27,556 elements for $n = 4$, together with the commitment to a redundancy function, for which some key axiomatic differences can only be sidestepped in the special case of two Gaussian variables \cite{Barrett2015, Kay.Ince2018}. Even in that simplest case, though, noise correlations can lead to significant interpretational problems \cite{Varley2024b}, prompting even a recent revision of the $\Phi$-ID metric itself \cite{Mediano.etal2025}. In summary, while $\Phi$-ID and PID represent valuable frameworks to resolve entropy-based measures more fine-grained \textit{in principle} (as demonstrated with the TE example), significant mathematical and interpretational questions that accompany this approach \textit{in practice} warrant carefulness in interpreting their atoms and associated bit values \cite{Varley2023b}. 

\subsection{Causal Information: From Prediction to Intervention}

Every measure in this section, however finely it resolves informational dependencies, is computed from a system left to its own devices, and this sets a ceiling for causal inference that no amount of resolution can raise. Consider two neuronal populations driven by a common third with slightly different conduction delays. The earlier will predict the later above and beyond its own history; TE will be non-zero, a directed link will be inferred, and $\Phi$-ID may even apportion the interaction into unique and synergistic parts. None of it distinguishes this arrangement from a genuine connection between the two, because the observational distribution can be identical in the two cases \cite{Mehler.Kording2020}. What separates a common cause from a direct one is not contained in any distribution we can passively observe; it resides in what the system would do under a perturbation, $P(X_1 \mid \mathrm{do}(X_2))$ \cite{Pearl2010}, which by construction no observational measure can access. To move from predicting the system to explaining it, we have to intervene on it and follow the time-asymmetric propagation of effects from there.

The higher-order vocabulary developed above carries over to this strictly larger space of distributions: interventional information can itself be decomposed into unique, redundant and synergistic contributions \citep{Jansma2025}, and causal emergence is precisely the case in which such a decomposition, evaluated under intervention, underwrites labels like `causal decoupling' or `downward causation' ($I_{\partial}(\{12\}\rightarrow\{12\})$ and $I_{\partial}(\{12\}\rightarrow\{1\})$, see previous section) with the interventional interpretation \citep{Mediano.etal2022} their labels already evoke. What changes is not the machinery behind the information measures but the question they answer, becoming about responses to manipulation rather than about prediction. 

The difficulty is that the ideal of surgically setting arbitrary nodes is rarely available in a behaving brain. Beyond the domain of simulated environments, we usually can neither so surgically interfere with every node nor can TE eliminate the causal directionality problem, as we have seen, unless we have some solid mechanistic priors. What is available to the experimenter is a weaker but attainable surrogate, the experimental contrast: rather than clamping an internal variable, the experimenter manipulates what enters the system and compares the resulting conditions. We usually cannot assign brain region states directly according to the randomised controlled trial (RCT) gold standard of causal inference but we \textit{can} randomise what stimulus features the brain responds to \cite{Schyns.etal2020, Yan.etal2023}. Feature-specific information transfer, for instance, isolates the component of a directed measure that concerns experimentally varied stimulus features rather than internal dynamics \citep{Ince.etal2015, Celotto.etal2023}, and stimulus-contrastive designs have been used to argue that cortical regions are jointly engaged in genuine stimulus processing above and beyond single-area encoding \citep{Gelens.etal2024, Varley.etal2022}. 

Every rung of the ladder we have climbed, from the undirected measures through the temporal decompositions to intervention itself, yields a signature of the system's behaviour; used carefully, they yield an informational `searchlight' into causally relevant information flows, but none of them unveils the generative mechanism directly. Reading mechanism and, ultimately, function off these signatures is a further inferential step, and one whose licence must be examined before any functional interpretation can be built on it.

\section{Functional Interpretations of Synergy and Redundancy}

\subsection{Distinguishing Between Higher-Order Mechanisms and Behaviours}

Intervention represents a crucial step towards elucidating mechanism from information but does not close that gap on its own. What every measure we have reviewed returns is a higher-order \textit{behaviour}: a statistical signature computed from the distribution of observed activity. What we would like to license are claims about higher-order \textit{mechanisms}: the generative structure that produced that distribution, its beyond-pairwise couplings, shared inputs, and update rules \citep{Rosas.etal2022}. Even under intervention, distinct mechanisms can produce identical behavioural signatures and, more strikingly, genuinely \textit{lower}-order mechanisms can generate higher-order behaviours \citep{Caprioglio.etal2026}.

A first reason the two come apart is that statistical behaviour depends on how we carve the system into variables and states, whereas the mechanism does not. The aggregation of a system's degrees of freedom into a description of measured variables and possible states, a choice of grain and basis that the experimenter makes, reshapes the redundancy--synergy balance we read out while leaving the underlying mechanism untouched \citep{Wheeler2018}. The point is concrete in neural data: Various recent studies on whole-brain dynamics have been conducted using highly coarse-grained fMRI data \citep{Luppi.etal2022,Luppi.etal2023,Luppi.etal2024a,Varley.etal2023a}, measuring the activity of $10^5-10^6$ neurons per spatial bin (voxel) \citep{Sadil.etal2022}, a massive difference from the circuit-level studies discussed that come closer to measuring $10^1-10^2$ neurons. This can lead to seemingly conflicting results from the same dataset for the same regions of interest (ROIs), if we choose, for instance, a finer-grained time resolution or a more microscopic measurement apparatus: Synergy at the microscopic circuit level is maximal at synaptic time scales \citep{Newman.etal2022, Luppi.etal2024c} but in macroscopically resolved fMRI recordings at much longer timescales \citep{Mediano.etal2022}. Similarly, we find in intracortical recordings stronger synergies within cortical columns than between them \citep{Nigam.etal2019} while the opposite relationship has been detected using fMRI \cite{Luppi.etal2022}. This serves as a reminder that inferences about synergy and redundancy emerge from within the measurement setup we choose.

None of this implies, though, that higher-order information signatures are merely epistemic. Beyond-pairwise structure \textit{can} be a genuine feature of a system's organisation, and a growing body of work aims to characterise such mechanisms directly rather than infer them from behaviour \citep{Rosas.etal2022, Liardi.etal2025, Caprioglio.etal2026}. As behaviour is observer-relative, the mechanism is what remains once the observer's choices are marginalised away: the generative process itself. A promising route to solving the inverse problem of uncovering higher-order mechanisms from behaviours is through their topology, representing beyond-pairwise structure as hypergraphs or simplicial complexes and studying their dynamics \citep{Millan.etal2025}. The connection to the information-theoretic behaviours reviewed here is beginning to be mapped, with synergy-dominated activity found to track the presence of higher-dimensional topological features such as cavities in the system's state space \citep{Varley.etal2025}. 

Higher-order mechanisms are easiest to hold apart from their behavioural shadow when they are stipulated by construction, as in the rule-based systems we turn to below, where the logical dependencies are known in advance, and the behaviour can be read against them. We begin where the gap is narrowest, with small networks whose generative structure is stipulated in advance, so that behaviour can be read against a known mechanism before we attempt the harder inverse problem in real neural data. We start off associating redundancy-synergy balance with (1) directed network motifs in simple Boolean systems, then add (2) temporal evolution to explore functional dynamics, before adding to a functional interpretation (3) the crucial element of their role in adaptation to demands external to the system measured.  

\begin{figure}[h]
    \centering
    \includegraphics[width=1\linewidth]{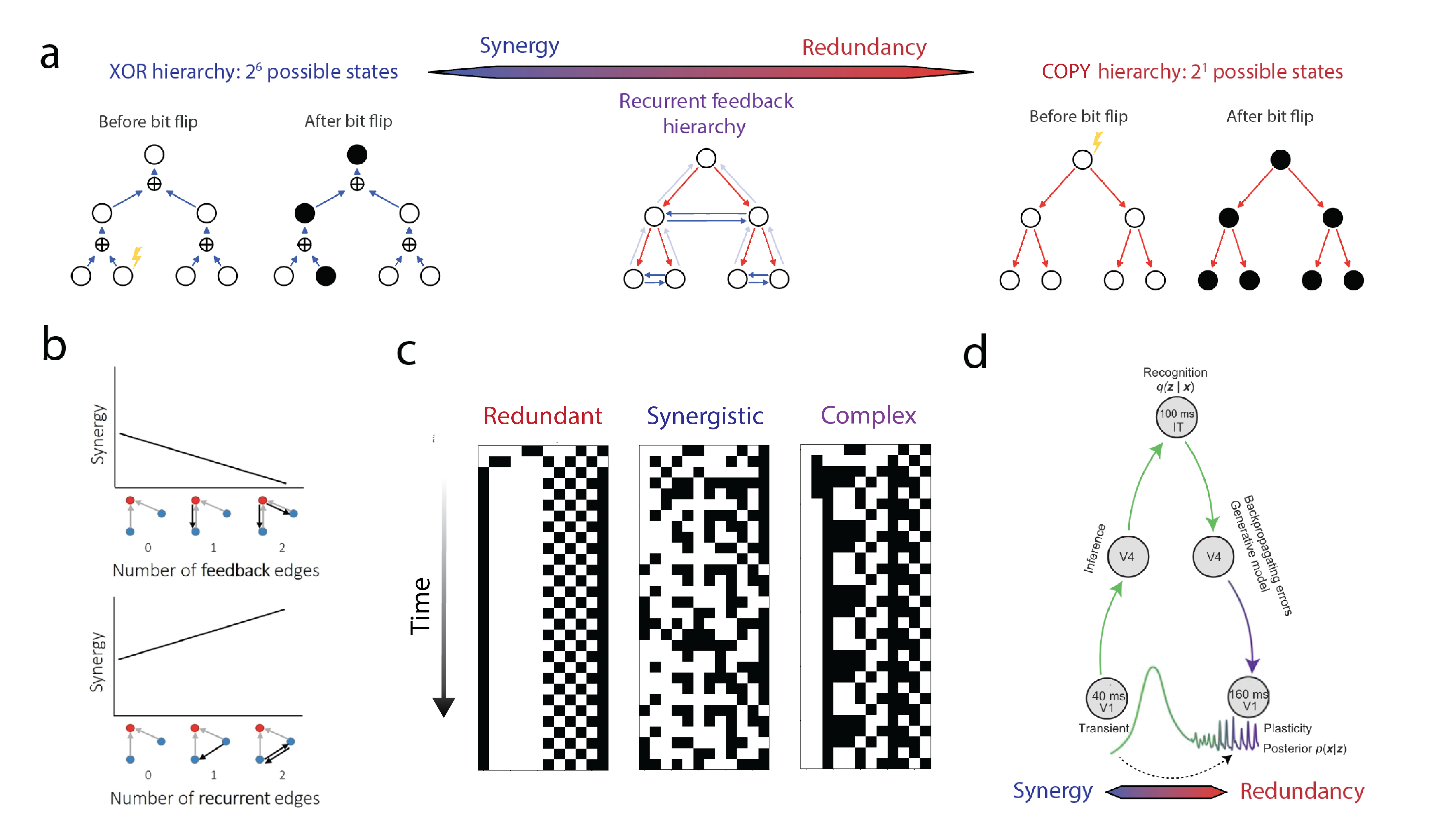}
        \caption{a) Simple network motifs of recurrent feedback hierarchies balance unconstrained bit flip cascades and lockstep evolution, supported by studies on dissociated cortical cultures, (b) showing that synergy is negatively associated with top-down feedback and positively with recurrence (from \cite{Sherrill.etal2021}). c) Boolean cellular automata were evolved to have maximal redundancy, synergy or TSE complexity. The networks with the highest TSE complexity seemed to balance the chaotic and periodic tendencies of the former two (from \cite{Varley.Bongard2024}). d) Synergy-dominated chaotic transients during recurrent visual feature integration are followed by redundant oscillations as low-level cortices are stabilized through top-down feedback (adapted from \cite{Vinck.etal2025}).}
    \label{applications}
\end{figure}

\subsection{Network Motifs of Redundancy-Synergy Balance and the Computation-Communication Tradeoff}

To inch closer towards characterizing functional purposes of synergy and redundancy, we consider two systems of seven variables with known directed dependencies that represent dual hierarchies in terms of redundancy and synergy (Figure \ref{applications}a). For the synergistic example, we find a feedforward hierarchy of XOR gates that recurrently integrates inputs at each level to inform the output at the next level. In this case, we find maximal sensitivity to inputs, because a bit flip in any of the input variables leads to a bit flip at the top of the hierarchy. Crucially, all the states of the `XOR hierarchy' jointly shape its state, where knowledge of all the other variables together is necessary to determine the output state. 

In contrast, a `broadcast hierarchy' of COPY operations allows all parts to move in synchrony, so that a bit flip in any variable (here placed at the top) determines the state of all other variables. In this way, the maximally redundant system is constrained to occupy only two of the $2^7$ possible joint states. The maximally synergistic system has a much higher degree of degeneracy in what joint states correspond to individual states, retaining half ($2^6$) of the allowed joint state repertoire. Thus, the parts move in synchronous lockstep for maximally redundant systems as the same state is effectively copied onto other parts. In contrast, parts move in complete context-dependence for maximally synergistic systems, where no context, short of considering all inputs together, defines a part.

The two prototypical directed motifs of information-sharing behind redundancy and synergy may serve complementary, dual purposes within a complex system: Redundant broadcasting of information replicates the same information from one part onto the rest for \textit{communication}, whereas synergistic integration creates information in one part from the rest for \textit{computation}. Several authors have noted that synergy is closely linked to computation, occurring when multiple inputs jointly influence the output of a third element \citep{Varley2024a,Timme.etal2014,Mediano.etal2022}. Although redundancy is most frequently associated with robustness \cite{Luppi.etal2022}, we find that its duality to synergistic computation can also be interpreted as a sign of large-scale broadcasting of information (communication) \cite{Sherrill.etal2021}. This can then underpin robust information availability across multiple variables (or neurons). 

Importantly, we find in synergy-redundancy balance a mirroring of the tradeoff between computation and communication: Communicating the same piece of information redundantly to many downstream parts stands in direct conflict with their capacity to compute their own state from the joint context. If a neuron strongly entrains its neighbors, the communication of its state diminishes their power to deliver their own contributions towards computation jointly. This is directly supported by a series of studies that investigated the network motifs associated with synergy and redundancy in organotypical cortical cultures at the circuit level \citep{Timme.etal2014,Timme.etal2016,Faber.etal2019,Sherrill.etal2020,Sherrill.etal2021,Newman.etal2022,Wibral.etal2017a,Shorten.etal2022}. One of the main takeaways is that synergy is associated with recurrent integration of inputs towards a joint upstream output, while feedback from target to sources decreases synergy \citep{Sherrill.etal2020,Sherrill.etal2021}. The authors attribute the negative correlation between feedback of the target neuron and synergy to the top-down signal that reduces the variance of upstream neurons, which they require to interact synergistically. This echoes the purely mathematical treatment with a functional explanation at the circuit level: Horizontal modulation of source neurons corresponds to increased synergy, while the top-down broadcasting of information reduces it.

\subsection{Information Dynamics: Synergistic Chaos and Redundant Oscillations}

So far, we have only related the logical structure in essentially atemporal probability distributions to cognitive function. However, cognition, at least in the brain, is typically assumed to unfold over time, which we must account for when investigating the impact of synergy and redundancy on \textit{dynamics}. Once more, we first take recourse to fully transparent Boolean systems to explore dynamics arising out of informational properties that are determined \textit{a priori}. Particularly insightful work in this domain has emerged from the study of synergy and redundancy in cellular automata (CAs) \citep{Wolfram1983}, where a one-dimensional array of cells iteratively updates its discrete state based on the states of its neighboring cells. This adds not only a notion of temporality (the system evolves over time steps) but also a notion of locality (only the local vicinity matters), moving it somewhat closer to real cognitive systems. 

Varley and Bongard \citep{Varley.Bongard2024} used a genetic algorithm to evolve Boolean networks' updating rules toward maximal and minimal O-information, and then investigated differences in the patterns these networks generate. While highly redundant networks quickly settled into a repetitive attractor state, highly synergistic ones showed average transient lengths comparable to those of randomly updating networks. Synergy acts here as a generator of novelty while redundancy constrains it. Thus, a single bit flip in the synergistic networks can set off interpenetrating cascades of bit flips down the line, reminiscent of chaotic dynamics.   

The chaotic dynamics associated with strongly negative O-information can be explained by the highly interdependent contributions of neighboring cells, in which single-bit flips are much more likely to also flip the output state than in redundant networks. The opposite holds for redundancy: Periodic, oscillatory dynamics, which are hardly sensitive to single-bit flips, robustly prevail. The less distributed logic operations of redundancy \citep{Orio.etal2023} can be of great advantage, though, when information needs to be reliably communicated with high fault tolerance. 

In line with the more repetitive oscillatory patterns found in redundancy-dominated CAs \cite{Varley.Bongard2024}, recent neuroscientific studies of local field potentials (LFPs) \cite{Olivares.etal2025, Roberts.etal2026} showed that narrowband oscillations are almost exclusively connected with redundancy-dominance. If we consider the fact that macroscopic oscillations picked up by LFPs are generally taken to arise from synchrony at the local circuit level \cite{Buzsaki.etal2012, Vinck.etal2023}, it is unsurprising that synchrony-based functional connectivity between these circuits reflects redundant higher-order ROI interactions \cite{Varley.etal2023}. As the brain accommodates surprising information \cite{Vinck.etal2025}, synergy-dominated transients within areas give way to redundancy-dominated oscillations \cite{Roberts.etal2025}, in which synergistic transients can be explained by recurrent feedback loops \cite{Gelens.etal2024, Blume.etal2026}. Large-scale narrowband oscillations may therefore index high-precision regimes in which predictive inference has reduced uncertainty, yielding low-dimensional, redundant dynamics suitable for local encoding and communication \cite{Schneider.etal2021, Roberts.etal2025}.       

It becomes evident that the brain needs to deploy redundant and synergistic information dynamics in concert to integrate information about the world without losing global coherence. Boolean networks with maximal TSE complexity showed this kind of balance: they exhibited longer transients than redundant networks but showed much clearer attractor dynamics than synergistic ones \citep{Varley.Bongard2024}. Likewise, the brain nests more irregular transients within slower oscillations. Through highly synergistic processing, the brain introduces the variability in outputs needed to flexibly respond to the richness of external states encountered in the environment, complying with Ashby's law of requisite variety \citep{Ashby1947, Boyd.etal2017}. In excess, however, this variability would not allow the state space to be sufficiently bounded to revolve around a set of homeostatic attractor states \citep{Friston.etal2014}. Redundancy can be thought of as ensuring canalization \citep{Waddington1942} of trajectories, converting variability in input sequences into coherence in output. What this output should be steered towards depends, though, entirely on the (task-)context external to the system.

\subsection{The Contextual Dimension: Connecting Measures to External Influences}

In this final section, we will turn to embedding the putative functional roles of synergy and redundancy within the contextual constraints under which cognition operates. Although Boolean systems provide an excellent ground for honing intuitions about the properties of redundancy and synergy, they are far too simple to capture the externally constrained functioning that evolution imposes on cognitive systems. We highlight three factors external to a cognitive system that significantly determine the role that synergy and redundancy end up playing within it: (1) task context, (2) resource constraints, and (3) exogenous noise. 

Understanding the functional purposes of redundancy and synergy helps interpret their use within a task context. Intracranial recordings from macaques' frontoparietal areas showed that both redundancy and synergy were significantly elevated during movement execution compared to the fixation, cue, and memory periods \cite{Varley.etal2022}. Redundancy was even more strongly elevated, though: As the cross-region interactions synchronise in a global response, this can perhaps be explained by a low-dimensional behavioural signal that is redundantly broadcasted across regions. Attentional state, similarly, reshapes the higher-order information dynamics of thalamocortical prediction-error learning \cite{Aijala.etal2026}. Mechanistic inferences about synergistic prediction error processing \cite{Gelens.etal2024, Olivares.etal2025} or long-range correlations \cite{Rajpal.etal2025} are only possible, though, if stimulus-evoked over spontaneous activity is taken into account. In this way, higher-order information can directly index which aspects of neural activity are recruited for function \cite{Park.etal2018, Zhan.etal2019}.

Synergistic integration of goal-relevant features plays an important computational role in any cognitive system that requires compressing its input-output mapping. Resource constraints, which disallow an injective (one-to-one) mapping of inputs onto outputs, have been shown in ANNs to be crucial to the emergence of synergy as a computational resource \citep{Ehrlich.etal2023a}. The algorithm learns to use the computational power of synergistic encoding with the limited resources it has \cite{Tax.etal2017}. Informational bottlenecks force ANNs towards synergistic dynamics, making them sensitive to information that must be considered jointly to be task-relevant \citep{Proca.etal2024}.

This efficient computation of outputs comes at the price of instability, though. Once noise corrupts the inputs, as it inevitably does in any natural setting, sufficient redundancy is needed to channel the chaos purely synergistic computation creates toward basins of attraction. When noise is introduced, the vulnerability of synergistic solutions is punished, forcing ANNs to redundantly overrepresent information \citep{Proca.etal2024}. Noise introduces not only system-side effects but also estimator-side effects: \cite{Schneidman.etal2003} showed how noise correlations can flip a code from synergistic to redundant (or vice versa) and that redundancy often increases as shared noise grows. More recently, \cite{Wollstadt.etal2023} demonstrated that noise first penalizes synergistic detectability. These contextual factors underscore once more the importance of ensuring that our models' readouts are carefully translated into functional interpretations.

\section{Conclusions and Future Directions}

Synergy and redundancy represent fundamental concepts in extending our understanding of cognition beyond bivariate interactions. In this review, we sought to provide intuition and guidance on how to connect mathematical tools to valid inferences about their complementary roles in cognitive functioning. We started by illustrating the promises and tradeoffs that come with climbing the rungs of higher-order information theory from entropy-based generalisations of mutual information to time-resolved interventional decompositions. Across these tiers, we hope to have made clear the various pitfalls that render a straightforward attribution of higher-order information to mechanistic processes fallacious and how to deal with them for careful inferences about cognitive function.

From purely mathematical considerations, it is evident that non-trivial interdependencies between many elements in a system require the amalgamation of synergy and redundancy. Redundancy collapses a system into rigid, low-dimensional synchrony, while pure synergy yields high-dimensional sensitivity that becomes functionally brittle. When translated from simple Boolean systems to neural circuits, we showed that the redundancy-synergy balance mirrors a computation-communication trade-off. The next challenge is to explain, using causal interventions, how this balance is realized across neural scales and timescales, and when it is adaptively shifted by context. This promises a more rigorous mapping of statistical measures onto the physical mechanisms that support cognitive functioning. 

Although the nascent community applying higher-order information theory to neuroscientific data seems to converge on putative functional interpretations of synergy and redundancy in the brain, many questions remain. 

The first issue that needs to be tackled more rigorously is the explicit connection of information dynamics across scales. Too few studies have, for instance, been conducted at the mesoscale of intracortical recordings (around $10^3$ neurons) to allow for an explanatory bridge between micro and macro. Explicit considerations of how microscopic and macroscopic measurements can be rendered mathematically consistent \cite{Varley.Hoel2022} and cognitively meaningful requires multiscale theoretical and experimental work, analogous to foundational work on the renormalisation group and associated experiments in physics. At the macroscale, the reach of these tools is expanding beyond the single brain altogether: higher-order information measures are now being brought to bear on inter-brain synergy in social coordination \citep{Chidichimo.etal2026}, extending the interpretive challenges catalogued here to interactions between nervous systems.

The second main obstacle to stronger, valid statements about the mechanistic roles of synergy and redundancy is the disentangling of their causal contributions from spurious detection during measurements. Due to their statistical degeneracy, maximally synergistic systems appear much closer to independent noise. Synergy coincides, by definition, with weak correlations between individual parts. Therefore, we are more likely to ascribe synergy to spurious correlations, which can easily appear from undersampling \cite{Gehlen.etal2024}. These spurious effects pertain in particular to macro-scale modalities such as EEG, MEG and fMRI, where the measured variables are approximately linear mixtures of many microscopic sources. Statistical biases induced by undersampling \citep{Gehlen.etal2024} or the lack of null-models \citep{Liardi.etal2025a} are just starting to be accounted for in this context. 

Ultimately, a fruitful direction towards a mechanistic, physically grounded interpretation of information atoms in brain data is to explicitly ground neuroscientific research in physical processes \citep{Friston.Frith2015, Liardi.etal2025} and their associated thermodynamic quantities \cite{Aguilera.etal2026}. Although foundational advances have come from researchers in statistical physics (e.g., \citep{James.etal2011,James.etal2016,James.Crutchfield2017}), concepts and tools out of this niche are only starting to enter mainstream computational neuroscience \citep{Luppi.etal2023b, Prokopenko.etal2024, Deco.Kringelbach2020}. Valuable questions include whether different forms of synergistic computation and redundant broadcasting carry distinguishable thermodynamic costs, and how the computation–communication trade-off can be connected to energy consumption or irreversibility in the brain. Ultimately, we advocate for embedding higher-order information measures within explicit physical generative models that predict how informational structure arises, evolves, and dissipates to drive cognitive complexity.

\section*{Competing interests statement}
The authors declare no competing interests.

\section*{Acknowledgments}
We would like to thank four anonymous reviewers for helpful comments on this manuscript. The neuron illustrations in Figure 1 were adapted from those created by John Chilton and obtained from SciDraw (10.5281/zenodo.3926009, 10.5281/zenodo.3926221, 10.5281/zenodo.3926143), licensed under CC-BY 4.0.

\section*{Authorship contributions}
Conceptualization: D.R. and A.C-J. Writing original draft: D.R. Editing: D.R., K.J.D., T.F.V., R.I. and A.C-J. Supervision: A.C-J.

\section*{Funding}
D.R. gratefully acknowledges financial support through the Amsterdam Science Talent Scholarship and the German Academic Scholarship Foundation. A.C-J. is supported by a Swedish Research Council project grant (VR; 2025-03245), a Research Council of Finland project grant (RCF; 375200), an ANID/FONDECYT Regular (1240899) and ANID/FONDECYT Regular (1251273) research grants.

\backmatter

\clearpage


\end{document}